\documentclass[aps,prl,twocolumn,showpacs,byrevtex]{revtex4}
\usepackage{graphicx} 
\usepackage{times} 
\usepackage{amsmath}

\newcommand\ie{i.e.,~} 
\newcommand\eg{e.g.~} 

\newcommand\etal{\emph{et al.}}

\begin{document}

\title{Density of near-extreme events}

\author{Sanjib Sabhapandit} 

\author{Satya N. Majumdar}

\affiliation{Laboratoire de Physique Th\'eorique et Mod\`eles Statistiques
  (UMR 8626 du CNRS), Universit\'e Paris-Sud, B\^atiment 100, 91405 Orsay
  Cedex, France}

\date{\today}

\begin{abstract}
  We provide a quantitative analysis of the phenomenon of crowding of
  near-extreme events by computing exactly the density of states (DOS) near
  the maximum of a set of independent and identically distributed random
  variables. We show that the mean DOS converges to three different
  limiting forms depending on whether the tail of the distribution of the
  random variables decays slower than, faster than, or as a pure
  exponential function. We argue that some of these results would remain
  valid even for certain {\em correlated} cases and verify it for power-law
  correlated stationary Gaussian sequences.  Satisfactory agreement is
  found between the near-maximum crowding in the summer temperature
  reconstruction data of western Siberia and the theoretical prediction.
\end{abstract}


\maketitle

Extreme value statistics (EVS)~\cite{EVT}, ---the statistics of the maximum
or the minimum value of a set of random observations,--- has seen a recent
resurgence of interests due to its applications found in diverse fields
such as physics~\cite{physics}, engineering~\cite{engineering}, computer
science~\cite{KM}, finance~\cite{finance}, hydrology~\cite{hydrology}, and
atmospheric sciences~\cite{climate}.  In particular, for independent and
identically distributed (i.i.d.)  observations from a common probability
density function (PDF) $p(X)$, the EVS is governed by one of the three well
known limit laws~\cite{EVT}, namely, (a)~Fr\'echet, (b)~Gumbel, or
(c)~Weibull, depending on whether the tail of $p(X)$ is, (a)~power-law,
(b)~faster than any power-law but unbounded, or (c)~bounded, respectively.
Recently, these same limiting laws have also been observed in a seemingly
different problem concerning the level density of a Bose gas and integer
partition problem~\cite{comtet}.

While EVS is very important, an equally important issue concerns the
near-extreme events~\cite{near-maxima}, ---\ie \emph{how many events occur
  with their values near the extreme}? In other words, whether the global
maximum (or minimum) value is very far from others (\emph{is it lonely at
  the top?}), or there are many other events whose values are close to the
maximum value.  This issue of the crowding of near-extreme events arises in
many problems.  For instance, in disordered systems, the low temperature
properties are governed by the spectral density function of the excited
states near the ground state.  In the study of weather and climate
extremes, an important question is: \emph{how often do extreme temperature
  events such as heat waves and cold waves occur?}  While for an insurance
company, it is very important to safeguard itself against excessively large
claims, it is equally or may be more important to guard itself from
unexpectedly high number of them.  In many of the optimization problems
finding the exact optimal solution is extremely hard and only practical
solutions available are the near-optimal ones~\cite{optimization}.  In
these situations, the prior knowledge about the crowding of the solutions
near the optimal one is very much desirable.

In this Letter, we study quantitatively the phenomenon of the crowding of
events near the extreme value for i.i.d. random variables, and find rather
rich and often universal behavior.  In general, the events that occur in
nature are correlated. However, when the correlations among them are not
very strong, then their EVS converges to that of the i.i.d. random
variables~\cite{berman}. This is why the limiting laws of EVS of the i.i.d.
random variables are very useful.  Here we consider i.i.d. random variables
in the similar spirit of the random-energy model~\cite{derrida} for
disordered systems, ---which despite its simplicity that the energy levels
are i.i.d. random variables, has been successful in capturing many
qualitative features of complex spin-glass systems.  Moreover, we provide
an example of a power-law correlated case, where the behavior of
near-extreme events converges to that of the i.i.d.  random variables.  In
addition, by comparing the near-maximum crowding in the reconstructed
summer temperature data of western Siberia against the prediction from the
i.i.d.  random variables, we find satisfactory agreement.

We start with a sequence of $N$ i.i.d. random observations $\{X_1, X_2,
\ldots X_N\}$, drawn from a common PDF $p(X)$. Let $X_{\max}$ be the maximum
of the sequence, ---\ie $X_{\max}=\max(X_1, X_2, \ldots X_N)$.  A natural
measure of the crowding of events near $X_{\max}$, is the density of states
(DOS) with respect to the maximum
\begin{equation}
  \rho(r,N) =\frac{1}{N} \sum_{\{X_i\neq X_{\max}\}}^{N-1}
  \delta\left[r-(X_{\max}-X_i) \right],
\label{DOS}
\end{equation}
where $r$ is measured from the maximum value, and we do not count
$X_{\max}$ itself, ---\ie $\int_0^\infty \rho(r,N)\, dr =1 -1/N$.  Clearly,
$\rho(r,N)$ fluctuates from one realization of the random sequence to
another, and one is interested in knowing whether its statistical
properties show any general limiting behavior, in the same sense, as one
finds for the EVS.  Note that, even though the random variables are
independent, the different terms in Eq.~(\ref{DOS}) become correlated
through their common maximum $X_{\max}$.

We find that the mean DOS $\overline{\rho(r,N)}$ displays rather rich
limiting behavior, as $N\rightarrow \infty$.  If the tail of the parent
distribution $p(X)$ of the random variables decays slower than a pure
exponential function, the behavior of $\overline{\rho(r,N)}$ is governed by
the corresponding extreme value distribution.  On the other hand, when the
tail of $p(X)$ is faster than a pure exponential, it is related to the
parent distribution itself.  In the borderline case when $p(X)$ has a pure
exponential tail, $\overline{\rho(r,N)}$ is entirely different.

To find $\overline{\rho(r,N)}$, first consider Eq.~(\ref{DOS}) for a given
value of the maximum at $X_{\max}=x$. Then the rest of the $(N-1)$
variables are distributed independently according to the common conditional
PDF $p_{\text{cond}} (X,x)=p(X)/\int_{-\infty}^x p(y)\,dy$. Hence the
conditional mean DOS, from Eq.~(\ref{DOS}), is
$\overline{\rho_{\text{cond}}(r,N,x)} =[(N-1)/N]\,p_{\text{cond}}(x-r,x)$.
For a set of $N$ i.i.d. random variables, the PDF of their maximum value
$X_{\max}=x$ is
\begin{equation}
  p_{\max}(x,N)= N p(x) \left[\int_{-\infty}^x p(y)\,dy\right]^{N-1}.
  \label{p_max}
\end{equation}
Thus, $\overline{\rho(r,N)}=\int_{-\infty}^\infty
\overline{\rho_{\text{cond}}(r,N,x)}\, p_{\max}(x,N)\,dx$. Upon substituting
the expressions for $\overline{\rho_{\text{cond}}(r,N,x)}$ and
$p_{\max}(x,N)$, a little algebra shows that
\begin{equation}
\overline{\rho(r,N)}=\int_{-\infty}^\infty p(x-r)\, p_{\max}(x,N-1)\, dx.
\label{g.1}
\end{equation}
This is the key result, which is valid for all $N$. We next analyze its
limiting behavior for large $N$.

For i.i.d. random variables, it is known that $p_{\max}(x)$ has a limiting
distribution~\cite{EVT}:
\begin{equation}
  b_N\, p_{\max}(x=a_N + b_N z,N)\xrightarrow{N\rightarrow \infty} f(z).
\label{limiting maximum distribution}
\end{equation}
The non-universal scale factors $a_N$ and $b_N$ depend explicitly on the
parent distribution $p(X)$ and $N$.  However, the scaling function $f(z)$
is universal and belongs to (a)~Fr\'echet, (b)~Gumbel, or (c)~Weibull,
depending only on the tail of $p(X)$.  For example, if $p(X) \sim
\exp(-X^\delta)$ for large $X$, then $a_N\sim (\ln N)^{1/\delta}$ and
$b_N\sim \delta^{-1} (\ln N)^{1/\delta-1}$ for large $N$, and the scaling
function is the universal Gumbel PDF $f(z)=\exp\left[-z-\exp(-z)\right]$.
Note that, as $N\rightarrow\infty$, for $\delta<1$, $b_N\rightarrow\infty$,
whereas $b_N\rightarrow 0$ for $\delta>1$.  In fact, this large $N$
behavior of $b_N$ is not restricted to only this specific tail of $p(X)$,
but is more generic: for any slower than $\exp(-X)$ tail of $p(X)$, as $N$
increases $b_N$ also increases, whereas for any faster than $\exp(-X)$
tail, $b_N$ decreases as $N$ increases.  This is indeed responsible for the
generic limiting behavior of $\overline{\rho(r,N)}$.

When $p(X)$ has a \emph{slower than exponential} tail, so that $b_N
\rightarrow \infty$ as $N\rightarrow\infty$, it is useful to make a change
of variable $x=a_N+b_N z$ in Eq.~(\ref{g.1}). Then one immediately realizes
that $p(b_N z +a_N-r)$ is highly localized, in the limit $N\rightarrow
\infty$, compared to $f(z)$, ---\ie $b_N p(b_N z +a_N-r) \rightarrow
\delta(z-[r-a_N]/b_N)$.  Therefore, in the scaling region of order $b_N$,
around $r=a_N$
\begin{equation}
\overline{\rho(r,N)}\xrightarrow{N\rightarrow\infty}
\frac{1}{b_N} f\left(\frac{r-a_N}{b_N} \right).
\label{dos slower than exp}
\end{equation}

On the other hand, if the tail of $p(X)$ is \emph{faster than exponential},
so that $b_N \rightarrow 0$ as $N\rightarrow\infty$, the PDF of the maximum
becomes highly localized near $x=a_N$, ---\ie $p_{\max}(x,N)\rightarrow
\delta(x-a_N)$.  Therefore, Eq.~(\ref{g.1}) yields
\begin{equation}
\overline{\rho(r,N)}\xrightarrow{N\rightarrow\infty} p(a_N-r).
\label{dos faster than exp}
\end{equation}

In EVS, the convergence towards the limiting distribution is usually very
slow~\cite{slow-convergence}. Therefore, it is instructive to check how
$\overline{\rho(r,N)}$ approaches the limiting form for large $N$. For this
purpose, now we consider explicit forms of $p(X)$, such that
$\overline{\rho(r,N)}$ can be computed to high accuracy for any given $N$
by numerically integrating Eq.~(\ref{g.1}), and also the explicit forms for
$a_N$ and $b_N$ as a function of $N$ can be obtained.  The mean number of
events close to the maximum, for a finite but large sample of size $N$, is
proportional to $\overline{\rho(0,N)}$.  In certain cases, $r=0$ is part of
the scaling function and $\overline{\rho(0,N)}$ can be obtained from the
scaling form of $\overline{\rho(r,N)}$ by putting $r=0$. However, sometimes
$r=0$ is not part of the scaling regime and $\overline{\rho(0,N)}$ has to
be computed separately from Eq.~(\ref{g.1}).  For simplicity, we consider
only positive random variables.

%%%%%%%%%%%%%%
\begin{figure*}
\includegraphics[width=7in]{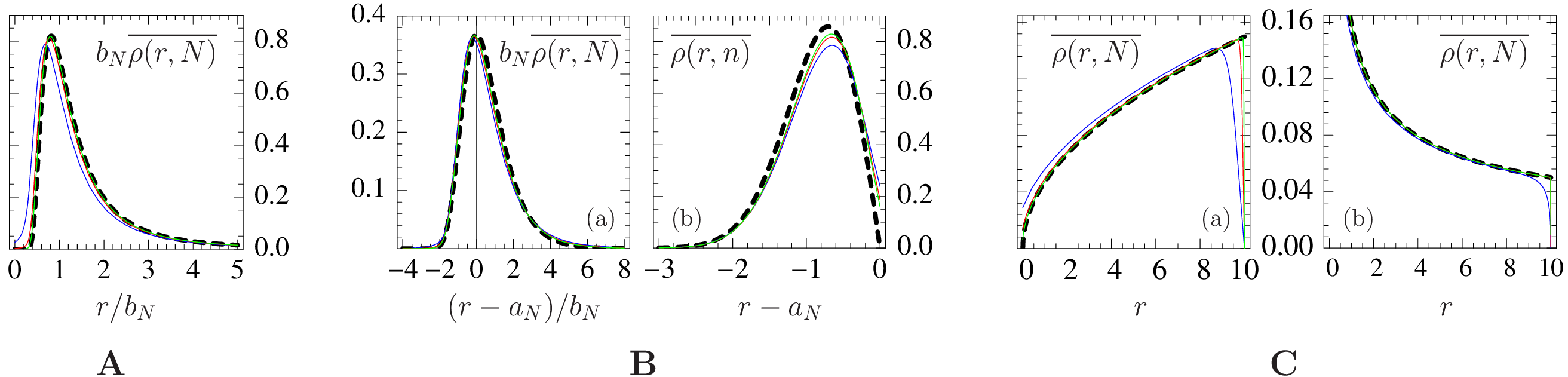}
\caption{\label{dosfigs} (Color online). {\bf A}: $\overline{\rho(r,N)}$
  for $N=10^2$ (blue), $10^3$ (red) and $10^4$ (green), for the power-low
  distribution $p(X)=\alpha \exp(-X^{-\alpha}) X^{-(1+\alpha)}$, with
  $\alpha=2$. The dashed (black) line plots the Fr\'echet distribution
  $f_1(r/b_N)$. {\bf B}: $\overline{\rho(r,N)}$ for exponential decay
  $p(X)=\delta X^{\delta-1} \exp(-X^\delta)$. (a) For $\delta=1/2$, with
  $N=10^3$ (blue), $10^5$ (red) and $10^7$ (green). The dashed (black) line
  plots the Gumbel distribution $f_2([r-a_N]/b_N)$. (b) For $\delta=2$,
  with $N=10^3$ (blue), $10^6$ (red) and $10^9$ (green).  The dashed
  (black) line plots $p(a_N-r)$. {\bf C}: $\overline{\rho(r,N)}$ for
  bounded distribution, $p(X)= \beta a^{-\beta} (a-X)^{\beta-1}$ for $X<a$
  and $p(X)=0$ for $X\ge a$, where $a=10$. (a) For $\beta=3/2$, with
  $N=10^2$ (blue), $10^3$ (red) and $10^4$ (green).  (b) For $\beta=1/2$,
  with $N=10$ (blue), $10^2$ (red) and $10^3$ (green).  The dashed (black)
  lines plot $p(a-r)$.}
\end{figure*}
%%%%%%%%%%%%%%

{\bf A}.~{\slshape\bfseries Power-law tail.}--- Consider $p(X)=\frac{\alpha
  \exp(-X^{-\alpha})}{ X^{1+\alpha}}$,
%$p(X)=\alpha \exp(-X^{-\alpha}) X^{-(1+\alpha)}$, 
where $\alpha>0$. In this case, $a_N=0$ and $b_N=N^{1/\alpha}$.  Therefore,
limiting $\overline{\rho(r,N)}$ is given by Eq.~(\ref{dos slower than
  exp}), with $f(z)$ belonging to the Fr\'echet class:
\begin{equation}
  f(z)\equiv f_1(z) =
  \frac{\alpha\exp\left[-z^{-\alpha}\right]}{z^{1+\alpha}}, \quad z\ge 0.
\label{frechet}
\end{equation}
Figure.~\ref{dosfigs}A compares this limiting form with the results obtained
from Eq.~(\ref{g.1}) by evaluating the integration numerically.  Here,
$r=0$ is away from the scaling regime. Thus, $\overline{\rho(0,N)}$ is
obtained directly from Eq.~(\ref{g.1}),
\begin{equation}
\overline{\rho(0,N)}\xrightarrow{N\rightarrow\infty} \frac{\alpha
\Gamma(2+1/\alpha)}{N^{1+1/\alpha}}.
\label{0-power}
\end{equation}

{\bf B}.~{\slshape\bfseries Faster than power-law, but unbounded tail.}---
Consider $p(X)=\delta X^{\delta-1} \exp(-X^\delta)$, where $\delta>0$. In
this case $a_N=(\ln N)^{1/\delta}$ and $b_N= \delta^{-1}(\ln N)^{1/\delta
  -1}$.  For very large and very small $r$, the large $N$ forms of the mean
DOS have same forms for all $\delta$, ---\ie $\overline{\rho(r,N)}\sim N
p(r)$ for $r\gg a_N$, and $\overline{\rho(r,N)}\approx p(a_N-r)$ for $r\ll
a_N$.  Thus, at $r=0$
\begin{equation}
  \overline{\rho(0,N)}\xrightarrow{N\rightarrow\infty} p(a_N)= 
  \frac{\delta}{N} (\ln  N)^{1-1/\delta},
\label{0-exponential}
\end{equation}
for all $\delta$.  However, the scaling behaviors of $\overline{\rho(r,N)}$
are very different for the three cases: $\delta<1$, $\delta=1$, and
$\delta>1$.

{\bfseries Case I: $\delta<1$}. As $N\rightarrow\infty$,
$b_N\rightarrow\infty$.  Therefore, in the scaling regime around $r=a_N$,
---which, however, becomes larger as $N$ increases, as $b_N$ becomes
larger--- the limiting $\overline{\rho(r,N)}$ is again given by
Eq.~(\ref{dos slower than exp}), but now $f(z)$ belongs to the Gumbel
class:
\begin{equation}
f(z)\equiv f_2(z)=\exp\left[-z-\exp(-z)\right].
\label{gumbel}
\end{equation}
Figure.~\ref{dosfigs}B~(a) compares the limiting form with the results
obtained from Eq.~(\ref{g.1}) by numerical integration.

{\bfseries Case II: $\delta=1$}. In this case $b_N=1$.  In this borderline
case neither of the limiting forms, ---\ie Eq.~(\ref{dos slower than
  exp})~or~(\ref{dos faster than exp}), are reached in the large $N$ limit.
Instead, we find a completely different behavior:
$\overline{\rho(r,N)}=g(r-a_N)$, where the scaling function
\begin{equation}
  g(z)=e^z \left[1-
    \left(1+e^{-z}\right) e^{-e^{-z}}  \right].
%    \left(1+e^{-z}\right) \exp\left(-e^{-z}\right)  \right].
\label{dos exp}
\end{equation}

{\bfseries Case III: $\delta >1$}. As $N\rightarrow\infty$, $b_N\rightarrow
0$. Thus, $\overline{\rho(r,N)}$ now converges to the other form given by
Eq.~(\ref{dos faster than exp}), which is compared in
Fig.~\ref{dosfigs}B~(b), with the results obtained from
Eq.~(\ref{g.1}) by evaluating the integration numerically.

{\bf C}.~{\slshape\bfseries Bounded tail.}--- Consider $p(X)= \beta
a^{-\beta} (a-X)^{\beta-1}$ for $0<X<a$, where $\beta >0$, and $p(X)=0$
otherwise. In this case, $a_N=a$ and $b_N=a N^{-1/\beta}$. Therefore, again
$\overline{\rho(r,N)}$ now converges to the other form given by
Eq.~(\ref{dos faster than exp}).  The comparison with Eq.~(\ref{g.1}) is
illustrated in Fig.~\ref{dosfigs}C.  Again, $N$ dependence of
$\overline{\rho(0,N)}$ for large $N$, does not follow from the limiting
$\overline{\rho(r,N)}$. This is obtained directly from Eq.~(\ref{g.1}),
\begin{equation}
  \overline{\rho(0,N)}\xrightarrow{N\rightarrow\infty} \frac{(\beta/a)
    \Gamma(2-1/\beta)}{N^{1-1/\beta}}, ~\text{for} ~\beta>1/2.
  \label{0-bounded}
\end{equation}

To summarize the explicit results: When the tail of $p(X)$ is either
power-law or bounded, the convergence of $\overline{\rho(r,N)}$ to the
respective limits given by Eqs.~(\ref{dos slower than exp}) and (\ref{dos
  faster than exp}) are fast, as can be seen from Figs.~\ref{dosfigs}A and
\ref{dosfigs}C respectively. However, in the intermediate situation ---\ie
when $p(X)$ decays faster than power-law but not bounded, --- the
convergence is slow, as can be seen from Figs.~\ref{dosfigs}B~(a) and
\ref{dosfigs}B~(b).  In other words, the more $p(X)$ deviates from
$\exp(-X)$ in either direction (slower and faster), $\overline{\rho(r,N)}$
converges more quickly (with increasing $N$) to its limiting form.  As $N$
increases, the mean number of events close to the maximum, which is
proportional to $\overline{\rho(0,N)}$, decreases faster for $p(X)$ with a
broader tail [cf. Eqs. (\ref{0-power}), (\ref{0-exponential}) and
(\ref{0-bounded})].  This is also evident from the small $r$ behavior of
$\overline{\rho(r,N)}$ in the scaling regime, ---\ie from the peak to the
left in Figs.  \ref{dosfigs}A and \ref{dosfigs}B~(a): For $p(X)$ with a
power-law tail, $\overline{\rho(r,N)}$ has an essential singular behavior
$\exp(-N/r^\alpha)$ for small $r$ [cf.  Eq.~(\ref{frechet})], and for a
stretched-exponential tail ({\bf B} with $\delta<1$), as $r$ decreases from
$a_N$ in the scaling regime $\overline{\rho(r,N)}$ decreases
super-exponentially $\exp(-\exp([a_N-r]/b_N))$ [cf. Eq.~(\ref{gumbel})]. On
the contrary, for $p(X)$ having faster than $\exp(-X)$ tail, there is
crowding near the maximum value ($r=0$) [Figs.  \ref{dosfigs}B~(b) and
\ref{dosfigs}C].

Another measure of the loneliness of the maximum is the gap between the
maximum and the next highest value.  Let $Q(\epsilon|N)$ be the PDF of the
gap being $\epsilon$. Clearly
\begin{equation}
  Q(\epsilon|N)= N \int_{-\infty}^\infty p(z+\epsilon)\, p_{\max}(z,N-1)\, dz.
\end{equation}
In particular, when $p(X)\sim \exp(-X^\delta)$ for large $X$, we find the
limiting form
\begin{equation}
  Q(\epsilon|N) \xrightarrow{N\rightarrow\infty} 
  \frac{1}{b_N} \exp(-\epsilon/b_N).
\end{equation}
Thus, the typical gap is of the order $b_N$, which increases (decreases) as
$N$ increases for $\delta<1$ ($\delta>1$), ---consistent with the results
obtained form the study of mean DOS.

\begin{figure}
  \includegraphics[width=3.375in]{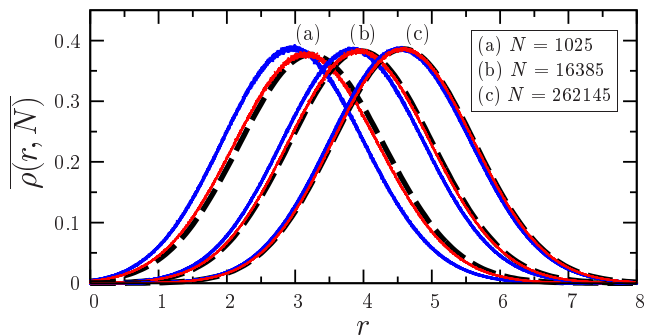}
  \caption{\label{gsp} (Color online). $\overline{\rho(r,N)}$ for stationary
    Gaussian random sequence with a correlator $C(n)=(1+n^2)^{-\gamma/2}$,
    where $\gamma=0.5$ (blue) and $\gamma=1$ (red) obtained from numerical
    simulation, and for Gaussian i.i.d. random variables (black dashed)
    obtained by numerical integration of Eq.~(\ref{g.1}). The three sets of
    curves (a), (b) and (c) correspond three different values of $N$.}
\end{figure}

So far, we have considered the case of i.i.d. random variables.  \emph{What
  would happen if the random variables are correlated?}  For short-ranged
correlation, one expects the results from i.i.d. random variables to hold.
However, for a stationary Gaussian sequence (SGS), this holds even for
long-range (\eg power-law) correlation.  More precisely, for SGS a rigorous
theorem~\cite{berman} states: if the correlator $C(n)\equiv\overline{X_i
  X_{i+n}}$ satisfies either $\lim_{n\rightarrow\infty} C(n) \ln n =0$ or
$\sum_{n=1}^\infty C^2(n) <\infty$, then the limiting distribution of the
maximum [cf.  Eq.~(\ref{limiting maximum distribution})] is Gumbel [cf.
Eq.~(\ref{gumbel})], and $a_N$ and $b_N$ are same as those in the case of
independent Gaussian random variables.  Based on this theorem, one
therefore predicts that $\overline{\rho(r,N)}$ for large $N$, should be
independent of the correlation function $C(n)$ and hence would be the same
as that of Gaussian i.i.d.  random variables.  We have indeed verified this
prediction for SGS's with a power-law correlation
$C(n)=(1+n^2)^{-\gamma/2}$, which are generated using numerical simulation.
We compute $\overline{\rho(r,N)}$ from these sequences for three different
values of $N$ and for each $N$ two different values of $\gamma$, and
compare with the one obtained by numerically integrating Eq.~(\ref{g.1})
for same $N$ and using $p(X)=\exp(-X^2/2)/\sqrt{2\pi}$, ---this is shown in
Fig.~\ref{gsp}. While for smaller $N$ [cf.  Fig.~\ref{gsp}~(a)] they
differ, for larger $N$ [cf. Fig.~\ref{gsp}~(c)] the difference becomes
unnoticeable.

%%%%%%%%%%%%%%%
%%%%%%%%%%%%%%

%%%%%%%%%%%%%%
\begin{figure}[b]
\includegraphics[width=3.375in]{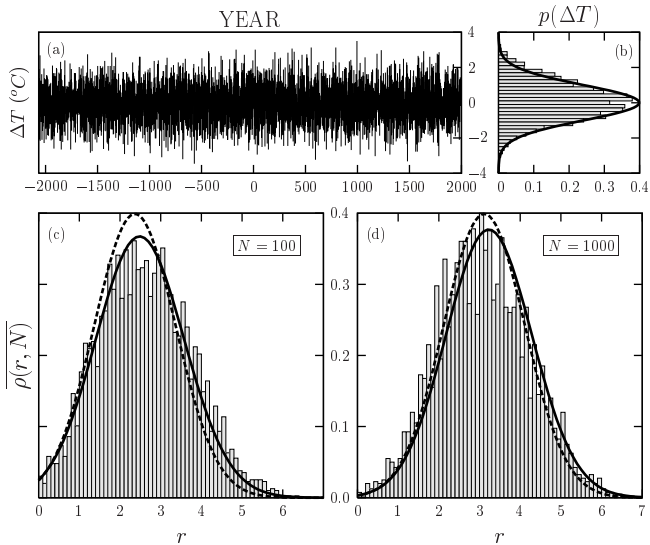}
\caption{\label{yamal} (a) Yamal peninsula June-July mean temperature
  anomaly ($\Delta T$) reconstruction series~\cite{yamal_data}. (b) The
  histogram plots the distribution $\Delta T$ of the data shown in (a). The
  solid line represents $p(\Delta T)=\exp(-\Delta T^2/2)/\sqrt{2\pi}$.  In
  (c) and (d), the histograms plot the mean DOS relative to the maximum
  (excluding the maximum), computed by dividing the data into blocks, with
  each block consists of $N$ years. Solid lines are calculated using the
  exact numerical integration in Eq.~(\ref{g.1}). The dashed lines represent
  $p(a_N-r)$, where $a_N=(2\ln N)^{1/2} - (2\ln N)^{-1/2} (\ln\ln N +\ln
  4\pi)/2$.}
\end{figure}
%%%%%%%%%%%%%%

\emph{How well do the mathematical results describe real data?}  That is
what we check last in this Letter, by comparing against the reconstructed
Yamal multimillennial summer temperature data by Hantemirov and
Shiyatov~\cite{yamal_data}. The reconstructed data-set consists of yearly
mean summer temperature anomalies ($\Delta T)$, of Yamal Peninsula of
western Siberia, relative to the mean of the full reconstructed series for
4000 years (2000 BC to AD 1996), which is shown in Fig.~\ref{yamal}~(a).
We divide the full time series into blocks of $N$ years, and for each
block: (I) find the maximum value of $\Delta T$, and then (II) with respect
to this maximum, compute $\rho(r,N)$ using Eq.~(\ref{DOS}).  Finally, we
find $\overline{\rho(r,N)}$, by taking average over all the blocks. The
histograms in Fig.~\ref{yamal}~(c) and (d) illustrate
$\overline{\rho(r,N)}$, computed by dividing the full series into
$40$-blocks with $100$ years of data in each block, and $4$-blocks with
$1000$ years of data in each block respectively.  Now to compare with our
results, we first compute the distribution of $\Delta T$ from the full time
series, which is illustrated in Fig.~\ref{yamal}~(b) by histogram, along
with the solid line given by the Gaussian distribution.  In
Fig.~\ref{yamal} (c) and (d), the solid lines are computed using the
Gaussian distribution from Eq.~(\ref{g.1}), by performing exact numerical
integration, with $N=100$ and $N=1000$ respectively. The dashed lines
correspond to the limiting form $p(a_N-r)$, obtained in Eq.~(\ref{dos
  faster than exp}) for large $N$. The agreements between them (dashed and
solid lines) are satisfactory.

% In summary, we have considered the DOS relative to the maximum of a set of
% $N$ i.i.d. random variables. We have shown that, the mean DOS converges
% different limits as $N\rightarrow\infty$, depending on whether the tail of
% the parent distribution of the random variables decay slower or faster than
% a pure exponential function.  We have compared our results against Yamal
% summer temperature data, and found satisfactory agreement.

We acknowledge the support of the Indo-French Centre for the Promotion of
Advanced Research under Project 3404-2.


\begin{thebibliography}{10}

\bibitem{EVT} R.A.~Fisher and L.H.C.~Tippet, Proc. Cambridge Philos. Soc.
  {\bf 24}, 180 (1928); %
  E.J.~Gumbel, \emph{Statistics of Extremes} (Columbia University Press, NY,
  1958); %
  J.~Galambos, \emph{The Asymptotic Theory of Extreme Order Statistics}
  (John Wiley \& Sons, NY, 1978).

\bibitem{physics} J.-P.~Bouchaud and M.~M\'ezard, J. Phys A {\bf 30}, 7997
  (1997); %
  D.S.~Dean and S.N.~Majumdar, Phys. Rev. E {\bf 64}, 046121 (2001); %
  G.~Gy\"orgyi, P.C.W.~Holdsworth, B.~Portelli, and Z.~R\'acz, Phys. Rev. E
  {\bf 68}, 056116 (2003); %
  S.N.~Majumdar and P.L.~Krapivsky, %Phys. Rev. E {\bf 62}, 7735 (2000); %
  Physica A {\bf 318}, 161 (2003); %
  J.F.~Eichner, J.W.~Kantelhardt, A.~Bunde, and S.~Havlin, Phys. Rev. E {\bf
    73}, 016130 (2006); %
  E.~Bertin and M.~Clusel, J. Phys. A {\bf 39}, 7607 (2006); %



\bibitem{engineering} A.N.~Norris, J. Mech. Materials Struct. {\bf 1}, 793
  (2006); %
  A.~Cazzani and M.~Rovati, Int. J. Solids Struct.  {\bf 42}, 5057
  (2005); %
  M.~Hayes and A.~Shuvalov, J. appl. mech. {\bf 65}, 786 (1998).



\bibitem{KM} S.N.~Majumdar and P.L.~Krapivsky, Phys. Rev. E {\bf 65},
  036127 (2002).


\bibitem{finance} P.~Embrechts, C.~Kl\"uppelberg, and T.~Mikosch,
  \emph{Modelling Extremal Events for Insurance and Finance} (Springer,
  Berlin, 1997).

\bibitem{hydrology} R.W.~Katz, M.B.~Parlange, and P.~Naveau, Advances in
  Water Resources {\bf 25}, 1287 (2002).

\bibitem{climate} D.R.~Easterling \etal, Science {\bf 289}, 2068 (2000); %
  S.~Redner and M.R.~Petersen, Phys. Rev. E {\bf 74}, 061114 (2006).


\bibitem{comtet} A.~Comtet, P.~Leboeuf, and S.N.~Majumdar, Phys. Rev. Lett.
  {\bf 98}, 070404 (2007).  

\bibitem{near-maxima} A.G.~Pakes and F.W.~Steutel, Aust. J. Stat. {\bf 39},
  179 (1997); %
  A.G.~Pakes and Y.~Li, Stat. Probab. Lett. {\bf 40}, 395 (1998).


\bibitem{optimization} D.S.~Dean, D.~Lancaster, and S.N.~Majumdar, Phys.
  Rev. E {\bf 72}, 026125 (2005).

\bibitem{berman} S.M.~Berman, Ann. Math. Stat. {\bf 35}, 502 (1964); %
  J.~Pickands, Trans. Am. Math. Soc. {\bf 145}, 75 (1969).


\bibitem{derrida} B.~Derrida, Phys. Rev. Lett. {\bf 45}, 79 (1980).

\bibitem{slow-convergence} G.~Gy\"orgyi, N.R.~Moloney, K.~Ozog\'any, and
  Z.~Ra\'cz, Phys. Rev. E {\bf 75}, 021123 (2007).



\bibitem{yamal_data} R.M.~Hantemirov and S.G.~Shiyatov, Holocene, {\bf 12},
  717 (2002). Data obtained from IGBP PAGES/WDC for Paleoclimatology,
  http://www.ncdc.noaa.gov/paleo/pubs/hantemirov2002/


\end{thebibliography}
\end{document}